\documentstyle[sprocl]{article}

\def\be{\begin{equation}}
\def\ee{\end{equation}}
\def\bea{\begin{eqnarray}}
\def\eea{\end{eqnarray}}
\begin{document}

\titlepage

\begin{flushright}
{COLBY 97-09\\}
{IUHET 372\\}
%{hep-ph/9711061\\}
{September 1997\\}
\end{flushright}
\vglue 1cm
	    
\begin{center}
{{\bf REVIVALS OF QUANTUM WAVE PACKETS\footnote[1]{\tenrm 
Talk presented by R.B. at the 
Fifth International Wigner Symposium,
Vienna, Austria, August, 1997}
\\}
\vglue 1.0cm
{Robert Bluhm$^1$, 
V. Alan Kosteleck\'y$^2$,
J.A. Porter$^3$,
and B. Tudose$^4$
\\} 
\bigskip
{\it $^1$Physics Department, Colby College\\}
\medskip
{\it Waterville, ME 04901, U.S.A.\\}
\vglue 0.3cm
{\it $^2$Physics Department, Indiana University\\}
\medskip
{\it Bloomington, IN 47405, U.S.A.\\}
\vglue 0.3cm
{\it $^3$Physics Department, Cornell University\\}
\medskip
{\it Ithaca, NY 14853, U.S.A.\\}
\vglue 0.3cm
{\it $^4$Physics Department, Northwestern University\\}
\medskip
{\it Evanston, IL 60208, U.S.A.\\}

\bigskip

\vglue 0.8cm
%{\tenrm ABSTRACT}
}
\vglue 0.3cm

\end{center}

{\rightskip=3pc\leftskip=3pc\noindent%\tenrm
We present a generic treatment of wave-packet revivals
for quantum-mechanical systems.
This treatment permits a classification of certain ideal 
revival types. 
For example,
wave packets for a particle in a 
one-dimensional box are 
shown to exhibit perfect revivals.
We also examine the revival structure of wave packets for
quantum systems with energies that depend on two quantum numbers.
Wave packets in these systems exhibit quantum beats in the
initial motion as well as new types of long-term revivals.
As an example, we consider the revival structure of a
particle in a two-dimensional box.
}

\vfill
\newpage

\title{REVIVALS OF QUANTUM WAVE PACKETS}

\author{ R. Bluhm }

\address{Physics Department, Colby College, Waterville, ME, 04901, USA}

\author{ V.A. Kosteleck\'y }

\address{Physics Department, Indiana University, Bloomington, IN, 04901, USA}

\author{ J.A. Porter }

\address{Physics Department, Cornell University, Ithaca, NY, 14853, USA}

\author{ B. Tudose }

\address{Physics Department, Northwestern University, Evanston, IL 60208, USA}

%%%%%%%%%%%%%%%%%%%%%%%%%%%%%%%%%%%%%%%%%%%%%%%%%%%%%%%%%%%%%%
% You may repeat \author \address as often as necessary      %
%%%%%%%%%%%%%%%%%%%%%%%%%%%%%%%%%%%%%%%%%%%%%%%%%%%%%%%%%%%%%%

\maketitle\abstracts{
We present a generic treatment of wave-packet revivals
for quantum-mechanical systems.
This treatment permits a classification of certain ideal 
revival types. 
For example,
wave packets for a particle in a 
one-dimensional box are 
shown to exhibit perfect revivals.
We also examine the revival structure of wave packets for
quantum systems with energies that depend on two quantum numbers.
Wave packets in these systems exhibit quantum beats in the
initial motion as well as new types of long-term revivals.
As an example, we consider the revival structure of a
particle in a two-dimensional box.
}
  
\section{Introduction}

The revival structure of quantum
wave packets is the subject of much current research 
in atomic, molecular, chemical, 
and condensed-matter physics.\cite{ajp}
One of the most studied examples of wave packets
is Rydberg wave packets.\cite{rwp}
When a Rydberg wave packet first forms,
its motion is periodic with the same classical 
period $T_{\rm cl}$ as a particle moving on
a keplerian ellipse.
Indeed,
the initial wave packet can be described as 
a type of squeezed state.\cite{ss}
However,
after several classical cycles, 
the wave packet collapses and a sequence of
fractional/full revivals and superrevivals commences.\cite{rwp,ap,sr}

An analysis of wave-packet revivals can be applied to a
variety of quantum systems other than Rydberg atoms.\cite{ajp,emp}
In fact,
a classification of different revival types is possible
based solely on the form of the energy $E_n$
for quantum systems with discrete energy levels.
This classification is described in the next section.
We then describe,
in the third section,
the revival structure of wave packets 
for quantum systems with energies that depend on
two quantum numbers.
We have proved that such systems can exhibit fractional revivals.\cite{revs2}
Examples of multi-dimensional systems that exhibit
fractional revivals are
a particle in a two-dimensional box,\cite{revs2}
Stark wave packets,\cite{stark}
and molecular wave packets.\cite{schinke}
  
\section{Classification of Types of Revivals}

The generic form of a quantum wave packet is a
superposition of energy eigenstates with energies $E_n$.
If the superposition is formed using a short laser pulse,
then the weighting of the energy states is strongly peaked
around a mean value $\bar n$.
This permits an expansion of the energy in a Taylor
series around $\bar n$.
The derivative terms in this expansion define 
time scales that govern the revival structure of
the wave packet:
$T_{\rm cl} = {2 \pi}/{E_{\bar n}^\prime}$,
$t_{\rm rev} = {- 4 \pi}/ E_{\bar n}^{\prime\prime}$,
and $t_{\rm sr} = {12 \pi} /{E_{\bar n}^{\prime\prime\prime}}$.

These definitions create three types of revival structure.\cite{ajp}
The first type occurs for quantum systems with energies $E_n$
that are linear in $n$.
For these systems,
only the classical period $T_{\rm cl}$ is defined,
and the motion of the wave packet is exactly periodic,
with no collapse or revivals.
An example of this type of revival structure
is a wave packet for a simple harmonic oscillator.
The second type of revival structure occurs for quantum
systems with energies that are quadratic in $n$.
For these systems,
the times $T_{\rm cl}$ and $t_{\rm rev}$ are defined,
but there is no superrevival time $t_{\rm sr}$.
An example of this type of system is a particle in a box.
As we have demonstrated,\cite{ajp}
the revivals of a wave packet for a particle in a
box are {\it perfect}. 
This is because there is no third-order time
scale $t_{\rm sr}$ that can modulate the motion
of the wave packet.
The fractional revivals therefore exhibit no distortion and are
reproduced exactly at mutiples of $t_{\rm rev}$.
It is this regularity of the fractional revivals that
accounts for some of the striking features seen more 
recently in  the ``quantum carpets''  described in 
this volume.\cite{schleich}
The third and most general type of revival structure occurs
for systems with energies that have more complicated dependence
on $n$.
Rydberg wave packets are an example of this class.
Their energies are $E_n = -1/2{n^2}$,
and therefore all three time scales $T_{\rm cl}$, $t_{\rm rev}$,
and $t_{\rm sr}$ are defined.
Wave packets in this type of system show both conventional
full/fractional revivals as well as full/fractional superrevivals.

\section{Multi-Dimensional Quantum Systems}

We have also examined the revival structure of wave packets
in quantum systems with energies that depend on two
quantum numbers.
For these multi-dimensional systems,
the revival structure is controlled by
two classical periods and three revival times.
These wave packets exhibit quantum beats in the
initial motion as well as new types of long-term revivals.
We have presented an analytical proof showing that fractional
revivals can occur in two-dimensional systems.\cite{revs2}
A condition that must hold is that the revival times
must be commensurate with each other.
This can usually be accomplished by tuning a
free parameter of the system.

As an example,
we have considered a
particle in a two-dimensional box.\cite{revs2}
Here,
the energy $E_{n_1 n_2}$ has separate quadratic dependence
on two quantum numbers.
For this reason,
there is a revival time associated with each
of the quantum numbers $n_1$ and $n_2$.
However,
the mixed revival time,
which depends on the derivative of $E_{n_1 n_2}$
with respect to both $n_1$ and $n_2$,
and all of the superrevival time scales vanish.
By selecting particular ratios of the length and width of the box,
different commensurabilities for the two revival times
can be chosen.
As in the case of a one-dimensional box,
the resulting full and fractional revivals are perfect,
and the pattern of revivals repeats after each full revival.

\section*{Acknowledgments}

This work is supported in part by the National
Science Foundation under grant number PHY-9503756.

\end{document}